\documentclass[prl,preprintnumbers,superscriptaddress,amsmath,amssymb,nofootinbib,twocolumn,floatfix]{revtex4-1}

\usepackage{setspace}
\usepackage{graphicx}
\usepackage{amsmath}  
\usepackage{ulem}

\usepackage{hyperref}
\usepackage{booktabs}    
\usepackage{siunitx}     
\usepackage{nameref}
\usepackage{multirow}
\usepackage{xcolor}

\newcommand{\sectionname}[1]{ \noindent{\bfseries #1}.---}

\newcommand{\be}{\begin{equation}}
\newcommand{\ee}{\end{equation}}

\usepackage{soul}

\numberwithin{equation}{section}

\begin{document}
\preprint{KCL-PH-TH-2025-38}

\title{Can machine learning improve the detectability and disentanglement of the gravitational-wave background?}

\author{Hugo Einsle}
\email{hugo.einsle@oca.eu}
\affiliation{Universiteit Antwerpen, Prinsstraat 13, 2000 Antwerpen, Belgium}\affiliation{Universit\'e C\^ote d'Azur, Observatoire de la C\^ote d'Azur, CNRS, Artemis, Nice 06300, France}
\author{Marie Anne Bizouard}
\email{marieanne.bizouard@oca.eu}
\affiliation{Universit\'e C\^ote d'Azur, Observatoire de la C\^ote d'Azur, CNRS, Artemis, Nice 06300, France}
\author{Tania Regimbau}
\email{regimbau@lapp.in2p3.fr}
\affiliation{LAPP, CNRS, 9 Chemin de Bellevue, 74941 Annecy-le-Vieux, France}
\author{Mairi Sakellariadou}
\email{mairi.sakellariadou@kcl.ac.uk}
\affiliation{Theoretical Particle Physics and Cosmology Group, \, Physics \, Department, \\ King's College London,  University of London,  Strand,  London, WC2R  2LS, UK}
\affiliation{Universit\'e C\^ote d'Azur, Observatoire de la C\^ote d'Azur, CNRS, Artemis, Nice 06300, France}
\author{Jishnu Suresh}
\email{jishnu.suresh@oca.eu}
\affiliation{Universit\'e C\^ote d'Azur, Observatoire de la C\^ote d'Azur, CNRS, Artemis, Nice 06300, France}

\newcommand{\TR}[1]{\textcolor{blue}{{#1}}}

\date{\today}
\begin{abstract}
Gravitational waves from compact binary coalescences and from early Universe processes are expected to form a gravitational-wave background. We employ a custom deep learning multi-scale multi-headed autoencoder architecture to isolate gravitational-wave background from detector noise, followed by a Markov chain Monte Carlo inference stage to separate the astrophysical and cosmological components. Analyzing $108$-day mock datasets representative of the first period of the fourth LIGO-Virgo-KAGRA  observing run, we show that we can detect with high confidence --- $\log_{10}$ noise Bayes factor larger than 3 --- a compact binary coalescence gravitational-wave background with an amplitude of $4.3^{+0.5}_{-0.4}\times10^{-9}$ at $f_{\rm ref}=25\,\mathrm{Hz}$, which is a factor $\sim5$ higher than the amplitude expected from compact binary sources. We also show that we can isolate a cosmological -- assumed flat spectrum -- gravitational-wave background as weak as $ 9.7^{+2.5}_{-2.4} \times 10^{-10}$ from the expected compact binary coalescence gravitational-wave background within simulated Gaussian noise mimicking the LIGO detectors sensitivity achieved in the fourth observing run. 
In blind-test comparisons with the standard \texttt{pygwb} pipeline, we  show that our method achieves more accurate amplitude and spectral-index recovery and enables the separation of astrophysical and cosmological background components.
\end{abstract}
\maketitle

\sectionname{Introduction}
The Universe is expected to be filled with a gravitational-wave background (GWB)~\cite{Christensen:2018iqi} of cosmological~\cite{Caprini:2018mtu}  or astrophysical~\cite {Regimbau:2011rp} origin. The former is generated by processes that took place in the early Universe, like topological defects, first-order phase transitions, an era of inflation. The latter is composed by a large number of  gravitational-wave (GW) signals originated from compact binary coalescences,  rotating or oscillating neutron stars, stellar core collapses.  
Detection of a GWB and separation of its components (cosmological from astrophysical) is one of the main targets of ground-based interferometeric gravitational-wave detectors~\cite{KAGRA:2013rdx}. 
GWB searches complement transient GW searches,
and can be used to constrain the star formation history of the Universe, as well as to test cosmological and high-energy physics models at energy scales not accessible by colliders~\cite{LIGOScientific:2025kry}.

We focus on current terrestrial gravitational-wave detectors, namely the LIGO-Virgo-KAGRA (LVK) network, which has recently completed its 4th observing run. LVK searches have already discovered over 390 compact binary mergers~\cite{LIGOScientific:2026wfs}, whilst one of their main targets remains the detection of the gravitational-wave background~\cite{LIGOScientific:2025bgj}.

Current GWB searches are based on cross-correlating calibrated strains from two or more widely-separated detectors~\cite{1999PhRvD..59j2001A}.
In the absence of correlated noise sources between the detectors, the total observation time is the only limiting factor of the search. Unlike for transient gravitational-wave signals, emitted
for instance from compact binary coalescences, for
a GWB search long integration times are required since the signal is much weaker than the intrinsic detector noise. In the presence of correlated noise between the detectors, one must accurately estimate and separate the relative strengths of the correlated noise and the GWB.
Presently performed GWB searches can account for the existence of correlated noise sources (e.g., Schumann resonances~\cite{Meyers:2020qrb}) between detectors.

Once a GWB is successfully detected, there will be the challenge of identifying the sources that contribute to it.
The development of the appropriate methodology for the separation of the astrophysical from the cosmological GWB is an ongoing effort. 
A Bayesian parameter estimation approach~\cite{Martinovic:2020hru} to statistically separate astrophysical from cosmological GWB contributions has shown that the
LIGO detectors, operating at design sensitivity~\cite{LIGOScientific:2014pky},
will not be able to distinguish among the two signals. 
A third-generation detector network  may potentially reveal a cosmological GWB provided one is able to reduce the astrophysical contribution to a residual background~\cite{2017PhRvL.118o1105R}, after the subtraction of sources detected individually    ~\cite{2020PhRvD.102f3009S,2020PhRvD.102f3009S,2023PhRvD.108f4040Z, Martinovic:2020hru}.
Remaining in the context of third-generation detectors, it was suggested~\cite{Zhong:2024dss} that applying a time-frequency domain notching procedure, or following
 a
Bayesian inference approach~\cite{Biscoveanu:2020gds} one may disentangle between the astrophysical and cosmological contributions.

We have recently proposed a novel hybrid approach~\cite{Einsle:2025xsh}, combining a multi-scale multi-headed autoencoder (MSMHAutoencoder) with Bayesian inference to separate the astrophysical from the cosmological GWB. We demonstrated its capability to separate a faint GWB signal from dominant detector noise more rapidly than traditional cross-correlation techniques. Using simulated data, we have shown that the LIGO-Virgo-KAGRA (LVK) network operating at design sensitivity, will be able to detect a GWB from compact binary mergers with fractional energy density $\Omega_{\rm GW}(f) \sim 10^{-9}$ at 25 Hz. Moreover, in the presence of such an astrophysical GWB this hybrid approach we have proposed will be able to simultaneously measure a flat spectrum cosmological component as faint as $\sim 1.3\times 10^{-10}$ using 24 days of data.

In this paper, we consider mock datasets simulating
O4a LIGO Hanford and LIGO Livingston noise, to test the ability of our multi-scale
multi-headed autoencoder architecture to confidently detect
 a compact binary coalescence gravitational-wave
background. Subsequently, we investigate the capacity of our architecture in isolating a cosmological GWB from the expected compact binary coalescence GWB within an O4a-like dataset. 

\sectionname{Methods} 
\label{sec:methods}
The GWB is modeled as the superposition of two components: an astrophysical background from compact binary coalescences (CBCs) and a cosmological background. The astrophysical component is generated in the time domain by summing individual signals from a population of binary black holes BBH, binary neutron stars BNS, and neutron star-black hole systems NSBH, using waveforms consistent with general relativity (see \cite{2025arXiv250612237T} for a full description). This produces a characteristic energy density spectrum $\Omega_{\text{astro}}(f) = \Omega_{\alpha} (f/f_{\rm ref})^{\alpha}$ where $\alpha=2/3$ for frequencies up to $\sim$100 Hz, reflecting inspiral-phase emission and the full contribution of the source population. Above this frequency, the spectrum flattens due to the reduced contribution from high-mass and high-redshift systems, which merge at lower observed frequencies. In practice, we use this spectrum and scale it to vary the amplitude. 
The cosmological component is modeled in the frequency domain as a stationary, Gaussian stochastic signal with cross-detector correlations governed by the overlap reduction function, assuming a scale-invariant spectrum $\Omega_{\text{cosmo}}(f) = \Omega_0$. The GWB from cosmic string loops is found to be flat in the high-frequency regime~\cite{Auclair:2019wcv}. A flat spectrum~\cite{BICEP:2021xfz} is also expected for a primordial GWB~\cite{BICEP:2021xfz} sourced by an inflationary era.  

\begin{figure}[htbp]
    \centering
    \includegraphics[width=0.49\textwidth]{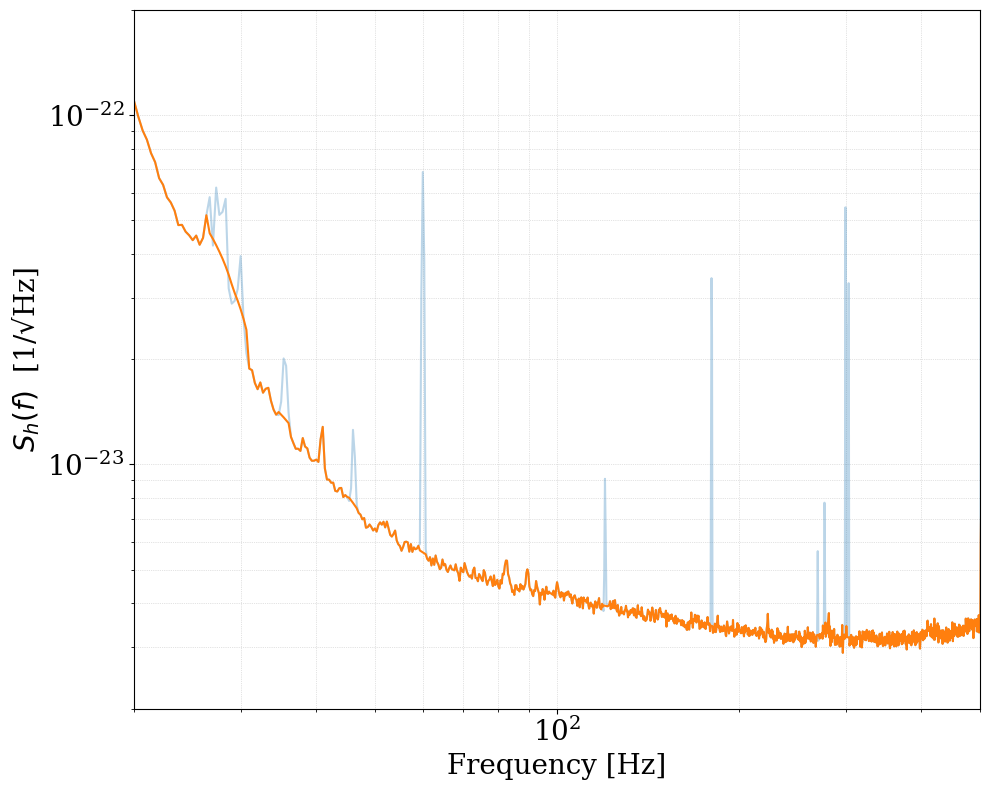}
    \caption{Representative amplitude spectral density curves for the LIGO Hanford detector during the O4a observing run~\cite{LIGO:2024kkz}. The original amplitude spectral density is shown in light blue, while the orange curve show the version without lines used to generate the idealized noise scenario for this study.}
    \label{fig:o4a_psds}
\end{figure}

In this study, we extend our previous analysis~\cite{Einsle:2025xsh} by generating two new mock datasets, each with a volume equivalent to the LIGO-Virgo-KAGRA O4a data used to search for an isotropic GWB~\cite{Abac:2025bgj} ($\sim$ 108 days). They are both based on the power spectral density (PSD) representative of the O4a noise for the two detectors~\cite{LIGO:2024kkz}. The first dataset uses the O4a PSD, while the second one uses a simplified PSD where spectral lines have been removed as shown in Figure~\ref{fig:o4a_psds}. Both datasets include the same injected GWB signals. The analysis presented here remains restricted to frequencies in the range (20-100) Hz. 
To separate the GWB from the detector noise we use the MSMHAutoencoder deep learning architecture developed in~\cite{Einsle:2025xsh}. This architecture is made to separate all signal components from the detector noise. It also includes elements to cope with spectral lines and non stationary features of the noise. The performance of MSMHAutoencoder to cope with spectral lines is tested by comparing the results with the two mock datasets. We perform independent training runs for each mock dataset using the \texttt{MSMHAutoencoder} architecture, following the methodology described in~\cite{Einsle:2025xsh}. To ensure a controlled comparison and prevent overfitting to specific noise conditions, we employ identical hyperparameters and data preprocessing across both dataset training instances.
Following the separation of the GWB signal from the detector noise by the \texttt{MSMHAutoencoder}, we characterize the recovered signal components via Markov chain Monte Carlo (MCMC) sampling and compute the Bayes factor $\mathcal{B}=\mathcal{Z}_{\rm S}/\mathcal{Z}_{\rm N}$ to quantify the evidence for a signal-plus-noise hypothesis against a noise-only hypothesis, where $\mathcal{Z}_{\rm S}$ is the evidence for the noise plus GWB hypothesis, and $\mathcal{Z}_{\rm N}$ is the evidence for the noise-only hypothesis. To further disentangle the background components, we calculate a second Bayes factor, $\mathcal{B}_{\rm cosmo}=\mathcal{Z}_{\rm CBC+cosmo}/\mathcal{Z}_{\rm CBC}$, comparing a model containing noise, a CBC background, and a cosmological background to a model containing noise and a CBC background only, where $\mathcal{Z}_{\rm CBC+cosmo}$ and $\mathcal{Z}_{\rm CBC}$ are the evidences for the respective hypotheses.


\begin{figure}[htbp]
    \centering
    \includegraphics[width=0.48\textwidth]{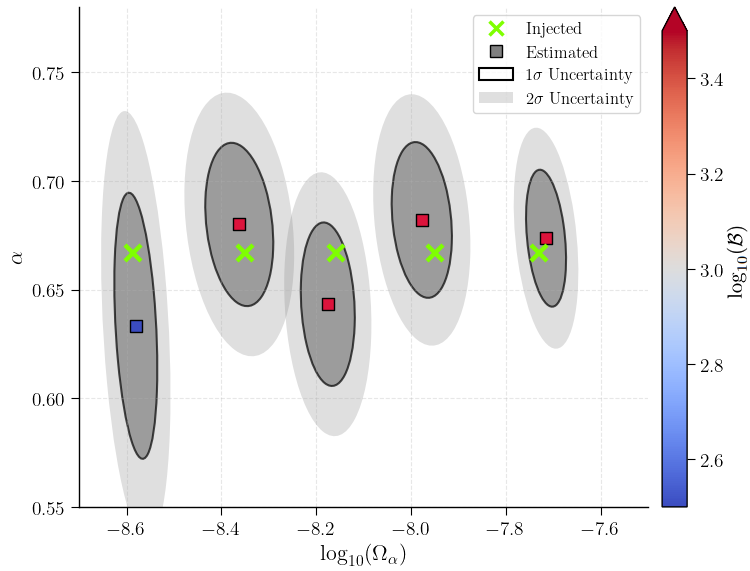} 
    \caption{Recovered median $\log_{10}(\Omega_{\rm \alpha})$ and $\alpha$ (square markers) shown inside $1\sigma$ and $2\sigma$ contours for five different amplitudes of the astrophysical GWB added to the O4a mock dataset. The $\log_{10}({\mathcal{B}})$ of each estimate is indicated by the color of the markers. The crosses represent the injected GWB spectrum parameters. 
    }
    \label{fig:sens_astro_alllines}
\end{figure}

\begin{figure}[htbp]
    \centering
    \includegraphics[width=0.48\textwidth]{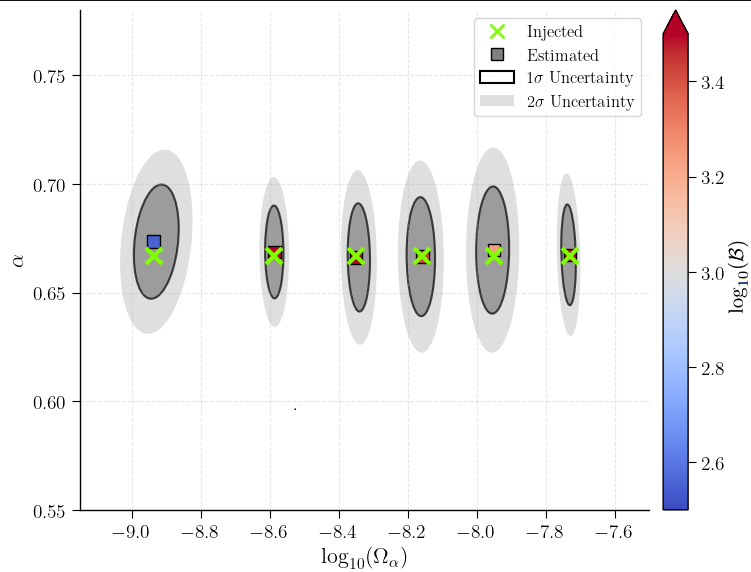} 
    \caption{Recovered median $\log_{10}(\Omega_{\rm \alpha})$ and $\alpha$ (square markers) shown inside $1\sigma$ and $2\sigma$ contours for six different amplitudes of the astrophysical GWB added to the O4a mock dataset for which lines have been removed. The $\log_{10}({\mathcal{B}})$ of each estimate is indicated by the color of the markers. The crosses represent the injected GWB spectrum parameters. 
    The non-uniformly varying scale of the ellipses between $\log_{10}(\Omega_\alpha)=-7.6$ and $\log_{10}(\Omega_\alpha)=-8.6$ is due to the curriculum learning focusing on the lowest amplitude signals.
    }
    \label{fig:sens_astro_nolines}
\end{figure}

\sectionname{Results}
\label{sec:results}
We first evaluate the capability of our method to detect an isolated astrophysical GWB. The detection threshold is defined as the amplitude $\Omega_{\alpha}$ at $f_{\text{ref}}=25$~Hz that yields $\log_{10} (\mathcal{B}) > 3$, corresponding to decisive evidence for a signal. We consider a uniform $\mathcal{U}$ prior for $\log_{10}(\Omega_{\alpha})$ and $\alpha$, given by   $\log_{10}(\Omega_{\alpha}) \in\mathcal{U}(-13, -7)$ 
and $\alpha \in \mathcal{U}(-5, 5)$, respectively, both wide enough to avoid possible biases. We vary the amplitude of $\Omega_\alpha$ from $2\times 10^{-8}$ down to $10^{-9}$ by steps of $\sim 0.25$ in $\log_{10}(\Omega_\alpha)$ for the GWB injected into the datasets.  In Figure~\ref{fig:sens_astro_alllines} we draw the credible-region contours of the recovered $\log_{10}(\Omega_\alpha)$ and $\alpha$ parameters of each injected astrophysical GWB. 
The faintest background amplitude is $\Omega_{\alpha} = 4.3^{+0.5}_{-0.4} \times 10^{-9}$ recovered for an injected CBC background of $\Omega_{\alpha}= 4.5 \times 10^{-9}$ with $\log_{10} ({\mathcal{B}}) = 3.3$. The $\alpha$ spectral index is well estimated within $1\sigma$ accuracy. Figure~\ref{fig:sens_astro_nolines} shows similar results with spectral lines being removed from the O4a data. The detection sensitivity improves to $\Omega_{\alpha} = 2.6^{+0.1}_{-0.1} \times 10^{-9}$ for an injected amplitude of $\Omega_{\alpha}= 2.6 \times 10^{-9}$, recovered with $\log_{10} ({\mathcal{B}}) = 3.7$. The ellipses are also slightly smaller than the ones for the O4a dataset containing spectral lines.
By keeping or suppressing each spectral line in the data, we conclude that the 60 Hz line is the one that  dominates the sensitivity loss.
It is worth noting that the MSMHAutoencoder architecture has  several deep convolutional neural networks (Inception blocks) that aim at suppressing the narrow spectral features. The suppression is not perfect and further improvements are required especially in the context of real data searches.

We then investigate the ability of our method to separate  cosmological and astrophysical GWB components by considering astrophysical injections set at the detection threshold ($\rm log_{10}(\mathcal{B})>3$) of the corresponding scenario. For each case, we superimpose a flat cosmological background and evaluate the minimum amplitude that can be identified with high confidence during O4a through the Bayes factor $\mathrm{\mathcal{B}_{cosmo}}$. We adopt for the cosmological background amplitude $\Omega_0$ an appropriately wide log-uniform prior, $\log_{10}(\Omega_{0}) \in \mathcal{U}(-13, -7)$.

\begin{figure}[htbp]
    \centering
    \includegraphics[width=0.48\textwidth]{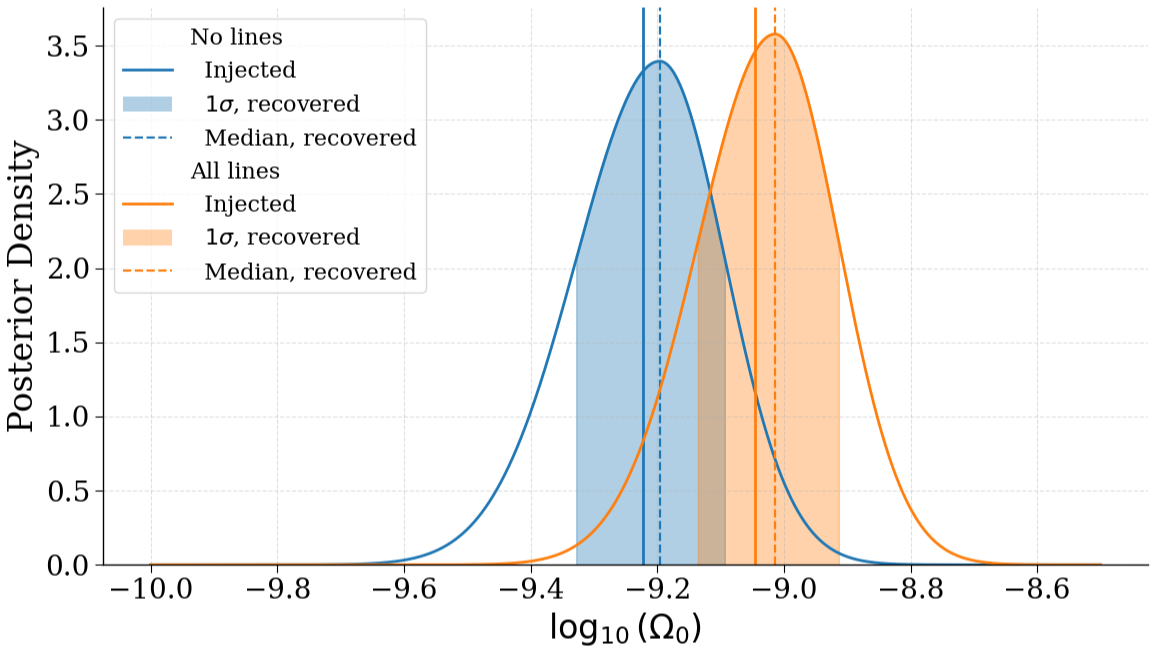} 
    \caption{Posterior distributions of the cosmological GWB amplitude $\log_{10}(\Omega_{0})$ recovered in the two O4a mock datasets in which a CBC GWB with an amplitude of $\Omega_\alpha = 4.5 \times 10^{-9}$ for the dataset with lines, and an amplitude of $\Omega_\alpha = 2.6 \times 10^{-9}$ for the dataset where spectral lines were removed, and $\alpha = 2/3$ has been injected. The blue posterior corresponds to the case where spectral lines have been removed from the data, while the orange posterior is obtained for the more realistic O4a-like noise dataset. The shaded regions indicate the $\pm 1\sigma$ credible intervals. The vertical solid lines mark the injected amplitudes $\log_{10}(\Omega_0)$, and the dashed lines show the recovered median values. In the no-lines case, a cosmological background injected at $\Omega_0 = 6 \times 10^{-10}$ is recovered as $\Omega_0 = 6.4^{+1.7}_{-1.7}\times 10^{-10}$, with $\log_{10}(\mathrm{\mathcal{B}_{cosmo}})=3.7$. In the more realistic noise case, a cosmological background injected at $\Omega_0 = 9 \times 10^{-10}$ is recovered as $\Omega_0 = 9.7^{+2.5}_{-2.4}\times 10^{-10}$, with $\log_{10}(\mathrm{\mathcal{B}_{cosmo}})=3.3$.}
    \label{fig:sens_cosmo_nolines}
\end{figure}

We report in Figure~\ref{fig:sens_cosmo_nolines} the posterior distribution of $\log_{10}(\Omega_{0})$ for both O4a mock datasets. We show that in the absence of spectral lines we can recover a cosmological background at $\Omega_0 = 6.4^{+1.7}_{-1.7}\times 10^{-10}$ injected as faint as $\Omega_0 = 6 \times 10^{-10}$, with $\log_{10}(\mathrm{\mathcal{B}_{cosmo}}) = 3.7$. In the more challenging realistic noise scenario, we can recover a cosmological background of $\Omega_0 = 9.7^{+2.5}_{-2.4} \times 10^{-10}$ injected at $\Omega_{0} = 9\times 10^{-10}$, with $\log_{10}(\mathrm{\mathcal{B}_{cosmo}}) = 3.3$.

\sectionname{Discussion}

Considering mock datasets simulating O4a LIGO Hanford and LIGO Livingston noise, we characterise the search sensitivity of the MSMHAutoencoder to a CBC-like background  as the faintest injected amplitude at $f_{\mathrm{ref}}=25\,\mathrm{Hz}$ still yielding decisive evidence. For the more realistic noise, this detection threshold is $\Omega_{\alpha}=4.5\times10^{-9}$ (recovered as $\Omega_{\alpha} = 4.3^{+0.5}_{-0.4} \times 10^{-9}$) with $\log_{10}({\mathcal{B}})=3.3$.
When spectral lines are removed, the detection threshold improves to $\Omega_{\alpha}=2.6 \times10^{-9}$ (recovered as $\Omega_{\alpha} = 2.6^{+0.1}_{-0.1} \times 10^{-9}$) with $\log_{10}({\mathcal{B}})=3.7$.

In~\cite{Abac:2025bgj} the LVK collaboration reports the result of the search for such an astrophysical GWB in the O4a data. No signal has been found, and a 95\% credibility upper limit of $\Omega_{\alpha} < 2.0 \times 10^{-9}$ is obtained using a log-uniform prior on $\Omega_{\alpha}$ and a uniform prior on $\alpha$, similar to those adopted in our analysis. Note, however, that this value cannot be compared directly with our detection threshold as the LVK collaboration upper limit is the amplitude excluded at 95\% credibility in the absence of a signal, whereas our result is the smallest amplitude yielding decisive evidence for a detection with $\log_{10}({\rm {\mathcal B}})> $ 3. In the case of a weak signal-to-noise ratio (SNR) signal, Gaussian noise, and equal priors for noise plus signal and noise-only hypotheses, Bayes factor and SNR are related through $\log({\mathcal B}) \approx {\rm SNR}^2/2$~\cite{PhysRevD.67.122002}. We could then compare our results, obtained for the minimal value of $\Omega_{\alpha}$ at $25$ Hz detectable with a SNR of $3$, with the cross-correlation method~\cite{2013PhRvD..88l4032T}. For $108$ days of O4a-like data, we obtain $\Omega_{\alpha} = 1.0 \times 10^{-8}$. If we then assume that detecting a GWB with a SNR of 3 can be compared to our high confidence detection threshold $\log_{10}({\mathcal{B}}) > 3$, then our MSMHAutoencoder result is a factor 2 better than what the cross-correlation method would achieve within the O4a data. This is equivalent to a reduction by a factor $\sim 5$ of the observing time required by the cross-correlation method to achieve the same result as our MSMHAutoencoder method.

One of the motivations of this machine learning development is not only to improve the capability to detect a GWB in LVK data, but also to distinguish a cosmological GWB from the astrophysical components. o test this capability, we conducted a simulated search for a cosmological GWB, assuming a flat spectrum, in 108 days of data containing a CBC GWB of amplitude corresponding to $\log_{10}({\mathcal B})=3$.
Considering the line-free dataset, which provides the best sensitivity, the MSMHAutoencoder significantly disentangles a cosmological background injected as faint as $\Omega_0 = 6 \times 10^{-10}$, recovered as $\Omega_0 = 6.4^{+1.7}_{-1.7}\times 10^{-10}$ with $\log_{10}(\mathrm{{\mathcal B}_{cosmo}})=3.7$, about a factor $1.5$ below the CBC GWB amplitude present in the data. For the more realistic O4a noise (including lines), the minimum detectable cosmological amplitude rises to $\Omega_0 = 9 \times 10^{-10}$, recovered as $\Omega_0 = 9.7^{+2.5}_{-2.4}\times 10^{-10}$ with $\log_{10}(\mathrm{{\mathcal B}_{cosmo}})=3.3$.

These results are consistent with our previous design-sensitivity study, which achieved \(\Omega_{\alpha}\simeq10^{-9}\) at 25\,Hz with \(\log_{10}({\mathcal B})> 3\) and a simultaneous detection of a cosmological component at \(\Omega_0\simeq1.3\times10^{-10}\) using about $27$ days of testing data. The difference is consistent with the factor $\sim 3$ between O4a and the expected LIGO design sensitivity. While increasing the amount of data reduces the variance of the MSMHAutoencoder output, the overall performance remains dataset-dependent. Specifically, the O4a data comprising spectral lines presents more complex features that degrade the reliability of the inference compared to the idealized noise used in the design-sensitivity study. Consequently, the gains from increased integration time are partially offset by the higher complexity of the instrumental background.

To further validate the results of our approach, we have also performed a blind-test analysis comparing the predictions of our DeepGWB pipeline with those obtained with the standard \texttt{pygwb} analysis~\cite{Renzini_2023}.
The test is described in the Appendix. We considered a GWB model with one or two components. The inferred parameters are reported in Tables~\ref{tab:model1} and~\ref{tab:model2_comparison}.
In addition to its ability to disentangle multiple background components, which is not accessible with the standard cross-correlation analysis, \textsc{DeepGWB} provides a more accurate recovery of the background amplitude than \texttt{pygwb}, and a significantly improved estimation of the spectral index even for the largest amplitudes (dataset~1 and 5).
Importantly, \textsc{DeepGWB} successfully reconstructs gravitational-wave backgrounds in regimes where \texttt{pygwb} does not achieve a significant detection, as illustrated by dataset~4.
We also find that the total GWB amplitude is consistently recovered by \textsc{DeepGWB}, with typical uncertainties at the level of \(\sim 10^{-10}\) for the considered configurations. In cases where a combined background is injected (CBC with $\alpha = 2/3$ and cosmological with $\alpha = 0$), \textsc{DeepGWB} tends to overestimate the cosmological component while underestimating the CBC contribution, although the total amplitude remains robustly reconstructed. Finally, we have checked that this effect is not generated by the MSMHAutoencoder algorithm, but remains inherent to the Bayesian parameter estimation.

Applying our algorithm on real LVK data is our next goal and is mandatory to fully validate this new machine learning method. We expect a loss of sensitivity due to non stationary features in the data. The presence of lines will also probably affect the performance as we shown in this study: despite the use of specific neural networks to well recognize the narrow spectral features, the lines within the $(20-100)$ Hz frequency band are decreasing significantly the performance (between 30\% and 60\%). As they persist in the training dataset, the neural network tends to interpret them as part of the stationary noise rather than artifacts to remove, and they are only weakly constrained by the loss function.
Adopting a more sophisticated MSMHAutoencoder architecture that is more resilient to spectral lines, it is expected to overall perform better than the standard preprocessing that consists in notching the spectral lines~\cite{Abac:2025bgj} and thus suppressing as well part of the signal. This is our aim for a follow-up study.


\sectionname{Acknowledgments}
The authors are grateful for computational resources provided by the LIGO Laboratory and supported by the National Science Foundation Grants PHY-0757058 and PHY-0823459. MS acknowledges support from the
Science and Technology Facility Council (STFC), UK, under the research grant ST/X000753/1. This work was supported by the French government through the France 2030 investment plan managed by the National Research Agency (ANR), as part of the Initiative of Excellence of Université Côte d’Azur under reference number ANR-15-IDEX-01.
This material is based upon work supported by NSF’s LIGO Laboratory which is a major facility fully funded by the National Science Foundation.




\bibliography{name}

\appendix

\section{Blind-test comparison between 
\texttt{pygwb} and \textsc{DeepGWB}}
\label{app:blindtest}

To further assess the performance of \textsc{DeepGWB}, we conducted a blind injection campaign whose objective was to compare the performance of \textsc{DeepGWB} with those of \texttt{pygwb}~\cite{Renzini_2023}, the isotropic stochastic-background analysis pipeline based on cross-correlation statistics developed within the LVK Collaboration.

The datasets consist of Gaussian detector noise mimicking O4a LIGO Hanford and LIGO Livingston noise~\cite{LIGO:2024kkz} together with injected GWB modeled as power laws. The injected signals include a CBC component with spectral index $\alpha = 2/3$, a cosmological component with $\alpha = 0$, or a combination of both. For each dataset, \textsc{DeepGWB} and \texttt{pygwb} were used to estimate the models' parameters as well as a detection significance as measured by the Bayes factors. The Bayesian priors are identical for both pipelines and are given in the main part of the article.
We first consider the recovery of a single GWB component either assuming the index spectral is fixed to 2/3, either inferring its value.
The recovered amplitude and spectral index values are given in Table~\ref{tab:model1} together with the parameters of the datasets. \textsc{DeepGWB} performs systematically better whatever the GWB amplitude as indicated by the $\log_{10}({\mathcal B})$ values. When the spectral index is a free parameter \textsc{DeepGWB} provides a more accurate recovery of the background amplitude than \texttt{pygwb}, and a significantly improved estimation of the spectral index even for the largest amplitudes (dataset~1 and~5). Even in these large amplitude cases, \texttt{pygwb} poorly infers the spectral index.
Importantly, \textsc{DeepGWB} successfully reconstructs GWBs in amplitude regimes where \texttt{pygwb} does not achieve a significant detection, as illustrated by dataset~4.
 stochasticWe also find that the total stochastic backgroundGWB amplitude is consistently recovered with  by \textsc{DeepGWB}, with typical uncertainties at the level of \(\sim 10^{-10}\) for the considered configurations. 

In Table~\ref{tab:model2_comparison}, we report the parameters of a two-component model inferred with \textsc{DeepGWB} only, as \texttt{pygwb} fails to converge regardless of the injected amplitudes.
In Table~\ref{tab:model2_comparison}, we report the parameters of a two-component model inferred with \textsc{DeepGWB} only, as \texttt{pygwb} was unable to recover the injected amplitude and spectral index even in the case of single component data set.

In cases where a two-components GWB is injected (CBC with $\alpha = 2/3$ and cosmological with $\alpha = 0$), \textsc{DeepGWB} tends to overestimate the cosmological component while underestimating the CBC contribution, although the total amplitude remains robustly reconstructed.

\begin{table*}
\caption{Blind-test comparison between \texttt{pygwb} and \textsc{DeepGWB} using simulated O4a-like LIGO Hanford and LIGO Livingston noise. The injected GWB includes a CBC component with spectral index $\alpha = 2/3$ (amplitude $\Omega_{\rm \alpha=2/3}$), a cosmological component with $\alpha = 0$ (amplitude $\Omega_{0}$), or a combination of both. All amplitudes are quoted in units of $10^{-9}$ at $f_{\rm ref}=25\,{\rm Hz}$. The uncertainties correspond to the 68.3\% credible intervals of the posterior distribution. Finally $\log_{10}({\mathcal B})$ gives an estimate of the significance of the estimate. When $\log_{10}({\mathcal B})<3$, parameters are poorly inferred. }
\label{tab:model1}
\begin{ruledtabular}
\begin{tabular}{c c c c  c c  c c  ccc  ccc}

\multicolumn{4}{c}{Injection} & \multicolumn{2}{c}{\textsc{DeepGWB}} & \multicolumn{2}{c}{\texttt{pygwb}} & \multicolumn{3}{c}{\textsc{DeepGWB}} & \multicolumn{3}{c}{\texttt{pygwb}}\\

\cline{2-4} \cline{5-6} \cline{7-8} \cline{9-11} \cline{12-14}

& $\Omega_{\alpha=2/3}^{\rm inj}$ & $\Omega_{0}^{\rm inj}$ & $\Omega_{\rm tot}^{\rm inj}$
& $\Omega_{\alpha=2/3}$ & $\log_{10}({\mathcal B})$
& $\Omega_{\alpha=2/3}$ & $\log_{10}({\mathcal B})$
& $\Omega_{\alpha}$ & $\alpha$ & $\log_{10}({\mathcal B})$
& $\Omega_{\alpha}$ & $\alpha$ & $\log_{10}({\mathcal B})$
\\
\hline
\\
1& 10.0 & 4.0 & 14.0 
 & $14.0^{+0.01}_{-0.01}$ & 33 
 & $15.05^{+1.62}_{-1.62}$ & 17.49
 & $15.20^{+0.11}_{-0.14}$ & $0.46^{+0.02}_{-0.02}$ & 31
 & $19^{+2.8}_{-2.7}$ & $-0.16^{+0.50}_{-0.55}$ & 7.46\\
\\
2&3.0 & 0.8 & 3.8
 & $3.83^{+0.04}_{-0.04}$ & 14 
 & $3.93^{+1.85}_{-2.39}$ & 0.84
 & $4.17^{+0.07}_{-0.07}$ & $0.45^{+0.04}_{-0.03}$ & 12 
 & $6.1^{+3.2}_{-3.9}$ & $-1.7^{+1.90}_{-1.92}$ & 0.37\\
\\
3&3.0 & 1.0 & 4.0 
 & $4.00^{+0.04}_{-0.04}$ & 15
 & $4.20^{+1.81}_{-2.20}$ & 0.96
 & $4.39^{+0.08}_{-0.06}$ & $0.43^{+0.03}_{-0.03}$ & 13
 & $6.5^{+3.2}_{-3.6}$ & $-1.6^{+1.85}_{-1.92}$ & 0.42
\\ \\
4&1.0 & - & 1.0  
& $1.01^{+0.02}_{-0.02}$ & 4.3
& $0.05^{+1.27}_{-0.05}$ & -0.05
&$1.08^{+0.04}_{-0.04}$ &$0.49^{+0.06}_{-0.08}$&3.1
&$0.05^{+241}_{-0.05}$&$-1.1^{+3.49}_{-2.71}$&-0.02
\\
\\
5&10.0 & 0.8 & 10.8 
 & $11.3^{+0.01}_{-0.01}$ & 73
 & $12.03^{+1.63}_{-1.63}$ & 10.80 
 & $11.70^{+0.11}_{-0.11}$ & $0.57^{+0.02}_{-0.02}$ & 71
 & $15^{+2.8}_{-2.8}$ & $-0.19^{+0.63}_{-0.70}$ & 4.52
\end{tabular}
\end{ruledtabular}
\end{table*}

\begin{table*}
\caption{
Blind-test separation of the CBC and cosmological background
components using \textsc{DeepGWB}.
The injected CBC and cosmological spectral indices are fixed to
\(\alpha_{\rm CBC}=2/3\) and \(\alpha_{\rm cosmo}=0\), respectively.
The results are reported both when the CBC spectral index is fixed to
\(\alpha=2/3\) and when it is inferred from the data.
All amplitudes are quoted in units of \(10^{-9}\) at
\(f_{\rm ref}=25\,{\rm Hz}\).
The uncertainties correspond to the 68.3\% credible intervals of the posterior distribution.
The Bayes factor
\({\mathcal B}_{\rm cosmo}
=\mathcal{Z}_{\rm CBC+cosmo}/\mathcal{Z}_{\rm CBC}\)
compares the evidences of the two-components model to the one of the CBC-only model.
}
\label{tab:model2_comparison}
\begin{ruledtabular}
\begin{tabular}{c c c c c c c c c c c}

& \multicolumn{3}{c}{Injection}
& \multicolumn{7}{c}{\textsc{DeepGWB}}
\\

\cline{2-4}\cline{5-11}

&
$\Omega_{\alpha=2/3}^{\rm inj}$
&
$\Omega_{0}^{\rm inj}$
&
$\Omega_{\rm tot}^{\rm inj}$
&
$\Omega_{\alpha=2/3}$
&
$\Omega_{0}$
&
$\log_{10}({\mathcal B}_{\rm cosmo})$
&
$\Omega_{\alpha}$
&
$\alpha$
&
$\Omega_{0}$
&
$\log_{10}({\mathcal B}_{\rm cosmo})$
\\

\hline
\\

1
& 10.0
& 4.0
& 14.0
& $9.84^{+0.35}_{-0.39}$
& $5.34^{+0.54}_{-0.40}$
& 7.0
& $9.78^{+0.45}_{-0.38}$
& $0.66^{+0.01}_{-0.01}$
& $5.40^{+0.54}_{-0.50}$
& 6.0
\\
\\

2
& 3.0
& 0.8
& 3.8
& $2.61^{+0.24}_{-0.15}$
& $1.58^{+0.25}_{-0.26}$
& 6.2
& $2.66^{+0.22}_{-0.19}$
& $0.66^{+0.01}_{-0.01}$
& $1.51^{+0.31}_{-0.23}$
& 5.7
\\
\\

3
& 3.0
& 1.0
& 4.0
& $2.62^{+0.22}_{-0.18}$
& $1.81^{+0.27}_{-0.27}$
& 5.3
& $2.67^{+0.22}_{-0.20}$
& $0.66^{+0.01}_{-0.01}$
& $1.72^{+0.31}_{-0.22}$
& 4.7
\\
\\

4
& 1.0
& -
& 1.0
& $0.76^{+0.06}_{-0.05}$
& $0.33^{+0.07}_{-0.06}$
& 2.9
& $0.78^{+0.06}_{-0.05}$
& $0.64^{+0.04}_{-0.05}$
& $0.32^{+0.08}_{-0.06}$
& 2.8
\\
\\

5
& 10.0
& 0.8
& 10.8
& $9.84^{+0.24}_{-0.24}$
& $1.84^{+0.35}_{-0.26}$
& 8.2
& $10.00^{+0.39}_{-0.32}$
& $0.65^{+0.03}_{-0.03}$
& $1.68^{+0.41}_{-0.32}$
& 7.3
\\

\end{tabular}
\end{ruledtabular}
\end{table*}

\clearpage
\onecolumngrid
\end{document}